\documentclass[aip, jap, amsmath, amssymb, reprint, longbibliography]{revtex4-1}

\usepackage{graphicx}
\usepackage{dcolumn}
\usepackage{bm}

\usepackage{hyperref}

\usepackage[utf8]{inputenc}
\usepackage[T1]{fontenc}
\usepackage{mathptmx}
\usepackage{etoolbox}

\makeatletter
\def\@email#1#2{%
 \endgroup
 \patchcmd{\titleblock@produce}
  {\frontmatter@RRAPformat}
  {\frontmatter@RRAPformat{\produce@RRAP{*#1\href{mailto:#2}{#2}}}\frontmatter@RRAPformat}
  {}{}
}%
\makeatother

\begin{document}

\title{Electron escape probability in high-efficiency photocathodes measured by reverse-injection photovoltage
}

\author{S. A. Rozhkov}
\email{rozhkovs@isp.nsc.ru}
\affiliation{Rzhanov Institute of Semiconductor Physics, Siberian Branch,
Russian Academy of Sciences, Novosibirsk 630090, Russia}
\affiliation{Novosibirsk State University, Novosibirsk 630090, Russia}

\author{D. A. Kustov}
\affiliation{Rzhanov Institute of Semiconductor Physics, Siberian Branch,
Russian Academy of Sciences, Novosibirsk 630090, Russia}

\author{V. V. Bakin}
\affiliation{Rzhanov Institute of Semiconductor Physics, Siberian Branch,
Russian Academy of Sciences, Novosibirsk 630090, Russia}

\author{V. S. Khoroshilov}
\affiliation{Rzhanov Institute of Semiconductor Physics, Siberian Branch,
Russian Academy of Sciences, Novosibirsk 630090, Russia}
\affiliation{Novosibirsk State University, Novosibirsk 630090, Russia}

\author{V. L. Alperovich}
\affiliation{Rzhanov Institute of Semiconductor Physics, Siberian Branch,
Russian Academy of Sciences, Novosibirsk 630090, Russia}
\affiliation{Novosibirsk State University, Novosibirsk 630090, Russia}

\author{O. E. Tereshchenko}
\affiliation{Rzhanov Institute of Semiconductor Physics, Siberian Branch,
Russian Academy of Sciences, Novosibirsk 630090, Russia}
\affiliation{Novosibirsk State University, Novosibirsk 630090, Russia}

\author{H. E. Scheibler}
\affiliation{Rzhanov Institute of Semiconductor Physics, Siberian Branch,
Russian Academy of Sciences, Novosibirsk 630090, Russia}
\affiliation{Novosibirsk State University, Novosibirsk 630090, Russia}
\affiliation{Novosibirsk State Technical University, Novosibirsk 630073, Russia}

\date{\today}

\begin{abstract}

The characterization of electron transfer through the emitting surface is of crucial importance for optimizing existing and developing new photocathodes. Here we propose and develop a method for the direct determination of the electron escape probability in high-efficiency semiconductor photocathodes. The proposed method is based on the variations in the surface photovoltage upon the injection of emitted photoelectrons back into a photocathode (``reverse injection''), which is induced by the polarity reversal of the external electric field. We demonstrate the method on \textit{p}-GaN(Cs,O) photocathodes with negative effective electron affinity by measuring the evolution of photoemission quantum efficiency upon the reverse injection of emitted electrons.

\end{abstract}
                              
\maketitle

\section{\label{sec:level1}Introduction}

\begin{figure*}[t]
\includegraphics[width=0.85\linewidth]{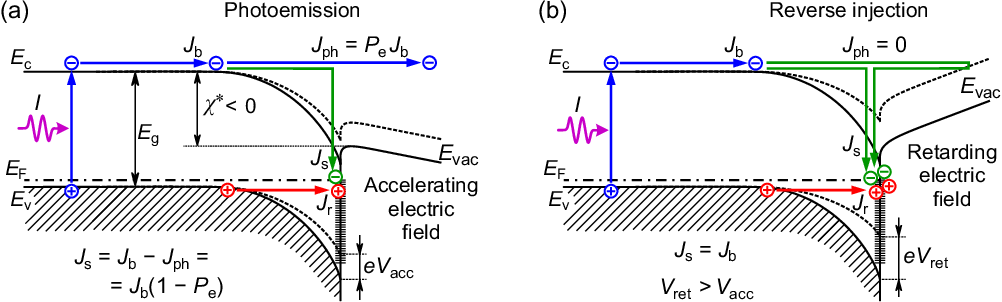}
\caption{\label{Fig.1}
(a) Formation of surface photovoltage in accelerating external electric field $V_{\text{acc}}$ during the photoemission from a photocathode with NEA. (b) Formation of photovoltage in retarding electric field $V_{\text{ret}}$ upon the reverse injection of emitted photoelectrons. The energy diagrams in the dark and at the above-band-gap illumination are shown by solid and dashed lines, respectively. $\chi^{*}$ is the effective electron affinity, $e$ is the electron charge modulus, $E_{\text{c}}$ and $E_{\text{v}}$ are the conduction band bottom and the valence band top, respectively, $E_{\text{g}}$ is the band gap, $E_{\text{F}}$ is the Fermi level, $E_{\text{vac}}$ is the vacuum level, $I$ is the optical power density, $J_{\text{b}}$ is the current of photoelectrons reaching the band bending region, $J_{\text{ph}}$ is the photoemission current, $P_{\text{e}}$ is the electron escape probability, $J_{\text{s}}$ is the surface recombination current, $J_{\text{r}}$ is the restoring current of equilibrium holes.
}
\end{figure*}

Semiconductor photocathodes with negative effective electron affinity (NEA) \cite{Bell1973} are widely used in practical applications and for scientific studies, including low-light imaging and single-photon detection \cite{Siegmund2003, Li2025}, as well as generation of spin-polarized \cite{Mamaev2008, Litvinenko2026} and ultra-cold electron beams \cite{Novotny2019, Rozhkov2026}. The widespread use of NEA photocathodes is due to the high values of the photoemission quantum efficiency (QE), reaching 50\% near the photoemission threshold \cite{Siegmund2003, Uchiyama2005}. According to the three-step photoemission model \cite{Vergara1990, Liu2026}: 
\begin{equation} \label{eq.1}
QE = P_{\text{ab}}P_{\text{tr}}P_{\text{e}},
\end{equation}
where $P_{\text{ab}}$, $P_{\text{tr}}$ and $P_{\text{e}}$ are the probabilities of photon absorption, photoelectron transfer to the emitting surface and electron escape into vacuum, respectively [see Fig.~\ref{Fig.1}(a)]. Further increase in QE requires optimizing all steps of the photoemission process. Absorption spectra and thus, $P_{\text{ab}}$ are known from the measured transmission and reflection spectra. The probability of the electron transfer to the emitting surface $P_{\text{tr}}$ depends on the recombination processes at the interfaces. Although these processes are well understood on the phenomenological level, the respective parameters, i.e., electron lifetimes, diffusion lengths and interface recombination velocities should be determined for each photocathode. The photoelectron escape probability $P_{\text{e}}$, which often restricts the quantum efficiency, depends in a complicated way on poorly known processes of electron transfer through the semiconductor-vacuum interface, cannot be calculated \textit{a priori} and should be determined experimentally.

The existing method for the $P_{\text{e}}$ determination is based on Eq.~\ref{eq.1}: QE is determined experimentally from the measured photocurrent and photon flux, while the values of $P_{\text{ab}}$ and $P_{\text{tr}}$ are calculated from the available data on the absorption spectra and recombination parameters. However, these data are typically not known with sufficient accuracy, and this leads to significant uncertainties in $P_{\text{e}}$ values \cite{Uchiyama2005, Liu2017}. Therefore, a direct method for determining the electron escape probability is needed to further optimize existing semiconductor photocathodes, as well as for the development of novel photocathodes \cite{Levenson2024, Rozhkov2024, Liu2026}.

In this work, we propose and develop a method for the direct determination of the electron escape probability by using the phenomenon of surface photovoltage (SPV) \cite{Kronik1999}. SPV consists in the decrease of surface band bending due to the screening of surface electric field by photogenerated electrons and holes. The decrease of band bending leads to lower photoemission quantum efficiency values because of decreasing the NEA modulus. Therefore, SPV is known to play a negative role in the operation of NEA photocathodes due to the effect of the surface charge limit \cite{Mulhollan2001}. The idea of the proposed method consists in measuring the SPV variation induced by the ``reverse injection'' of emitted electrons back into the photocathode. Here, the method is demonstrated by determining the electron escape probability in \textit{p}-GaN(Cs,O) NEA photocathodes.

\section{\label{sec:level1}Theoretical Framework}

The photoemission from a \textit{p}-type semiconductor photocathode with NEA at above-band-gap illumination with optical power density $I$ is shown in Fig.~\ref{Fig.1}(a). Upon reaching the emitting surface, photoelectrons either escape into vacuum or are captured in the surface band bending and recombine at the surface, since a return to the photocathode bulk is unlikely due to the scattering in the band bending region \cite{Liu2004, Pakhnevich2004}. Hence,
\begin{equation} \label{eq.2}
J_{\text{ph}} + J_{\text{s}} = J_{\text{b}} \propto P_{\text{ab}}P_{\text{tr}}I,
\end{equation} 
where $J_{\text{b}}$ is the current of photoelectrons that have reached the band bending region, $J_{\text{ph}}$ is the photoemission current and $J_{\text{s}}$ is the surface recombination current. The stationary non-equilibrium surface charge and, consequently, surface photovoltage $V$ are determined by the balance between $J_{\text{s}}$ and the restoring current of equilibrium holes $J_{\text{r}}(V)$ \cite{Rhoderick1978, Mulhollan2001}:
\begin{equation} \label{eq.3}
J_{\text{s}} = J_{\text{r}}(V) \approx J_{0} [ \exp(V/E_{0}) - 1],
\end{equation}
where $E_{0}$ and $J_{0}$ are the parameters, which depend on the mechanisms of hole transport to the surface through the band bending barrier.

The value of $J_{\text{s}}$ and, consequently, the magnitude of SPV depend on the direction of the external electric field. Upon applying the accelerating electric field [Fig.~\ref{Fig.1}(a)], the photoemission current reaches its maximum value of $P_{\text{e}}J_{\text{b}}$; so, $J_{\text{s}}$ and SPV reach their minimum values of $(1-P_{\text{e}})J_{\text{b}}$ and $V_{\text{acc}}$, respectively. Taking into account Eqs.~\ref{eq.2} and \ref{eq.3}, we obtain:
\begin{equation} \label{eq.4}
V_{\text{acc}} = E_{0} \ln [1 + (1 - P_{\text{e}}) J_{\text{b}} / J_{0}].
\end{equation}
On the contrary, the application of the retarding external electric field [Fig.~\ref{Fig.1}(b)] results in the reverse injection of emitted photoelectrons back to the photocathode surface. This leads to the increase in $J_{\text{s}}$ and SPV toward their maximum values $J_{\text{b}}$ and $V_{\text{ret}}$, respectively. Therefore,
\begin{equation} \label{eq.5}
V_{\text{ret}} = E_{0} \ln (1 + J_{\text{b}} / J_{0}).
\end{equation}
Thus, the surface photovoltages in accelerating and retarding external electric fields differ, and $P_{\text{e}}$ can be obtained by measuring the SPV variation upon the reverse injection of emitted photoelectrons.

It follows from Eqs.~\ref{eq.4} and \ref{eq.5} that, in the low optical power limit ($J_{\text{b}} \ll J_{0}$), the escape probability is determined by the ratio of surface photovoltages measured in accelerating and retarding external electric fields:
\begin{equation} \label{eq.6}
P_{\text{e}} \approx 1 - V_{\text{acc}} / V_{\text{ret}}.
\end{equation}
In the high optical power limit ($J_{\text{b}} \gg J_{0}$), the escape probability is determined by the difference in SPV for retarding and accelerating fields: 
\begin{equation} \label{eq.7}
\Delta V_{\text{ri}} = V_{\text{ret}} - V_{\text{acc}},
\end{equation}
\begin{equation} \label{eq.8}
P_{\text{e}} \approx 1 - \exp ( - \Delta V_{\text{ri}}/E_{0}).
\end{equation}

There are various experimental methods for studying SPV \cite{Kronik1999}. In vacuum a conventional technique consists in measuring the variations of the surface work function or, equivalently, contact potential difference (CPD). In large vacuum set-ups, CPD can be measured by the vibrating electrode technique (Kelvin probe) \cite{Kronik1999}. In vacuum photodiodes with parallel-plate geometry, the determination of CPD (and, hence, SPV) is possible by measuring photocurrent-voltage characteristics, as described below in section III and, in more detail, in Ref.\cite{Rozhkov2018}. However, measuring a photocurrent-voltage characteristic with a sufficient accuracy takes time, typically in the range from seconds to several minutes. In semiconductors, SPV kinetics usually proceeds on shorter timescales, and this limits the applicability of the conventional CPD technique. To overcome this limitation, we developed an alternative technique for the determination of the electron escape probability by measuring the quantum efficiency instead of the CPD variations. The technique is based on the fact that, to a good approximation, QE of a NEA photocathode is proportional to the change in the work function and, consequently, to the SPV \cite{Mulhollan2001, Rozhkov2018}:
\begin{equation} \label{eq.9}
QE \approx QE_{0} - AV,
\end{equation} 
where $QE_{0}$ is the quantum efficiency in the low optical power limit and $A$ is a proportionality coefficient. Taking into account Eqs.~\ref{eq.7} and \ref{eq.9}, we obtain that, upon the reverse injection of emitted photoelectrons, the decrease in the quantum efficiency  $\Delta QE_{\text{ri}}$ is proportional to the change in the surface photovoltage $\Delta V_{\text{ri}}$:
\begin{equation} \label{eq.10}
\Delta QE_{\text{ri}}  = -A \Delta V_{\text{ri}}.
\end{equation}
The value of $\Delta QE_{\text{ri}}$ can be obtained by measuring the relaxational kinetics of QE after the reversal of the external electric field polarity.

Despite the apparent simplicity of Eqs.~\ref{eq.6} and ~\ref{eq.9}, at low values of optical power, small values of SPV and, consequently, small variations in QE lead to large uncertainties in the resulting values of the electron escape probability. 
Therefore, the high optical power limit turned out to be more practical for the determination of $P_{\text{e}}$. Combining Eqs.~\ref{eq.8} and \ref{eq.10}, we obtain:
\begin{equation} \label{eq.11}
P_{\text{e}} \approx 1 - \exp [ \Delta QE_{\text{ri}} / (AE_{0})].
\end{equation}
Therefore, one can determine $P_{\text{e}}$ by measuring the dependence of QE on $I$, along with the QE kinetics following the reversal of the external electric field polarity. It should be noted that, in Eqs.~\ref{eq.4},~\ref{eq.6},~\ref{eq.8} and ~\ref{eq.11},  $P_{\text{e}}$ corresponds to the photocathode at photovoltage $V_{\text{acc}}(I)$. The value of escape probability in the low optical power limit $P_{\text{e, 0}}$ can be obtained by multiplying $P_{\text{e}}$ by the ratio $QE_{0}/QE(I)$.

\begin{figure}[t]
\includegraphics[width=0.88\linewidth]{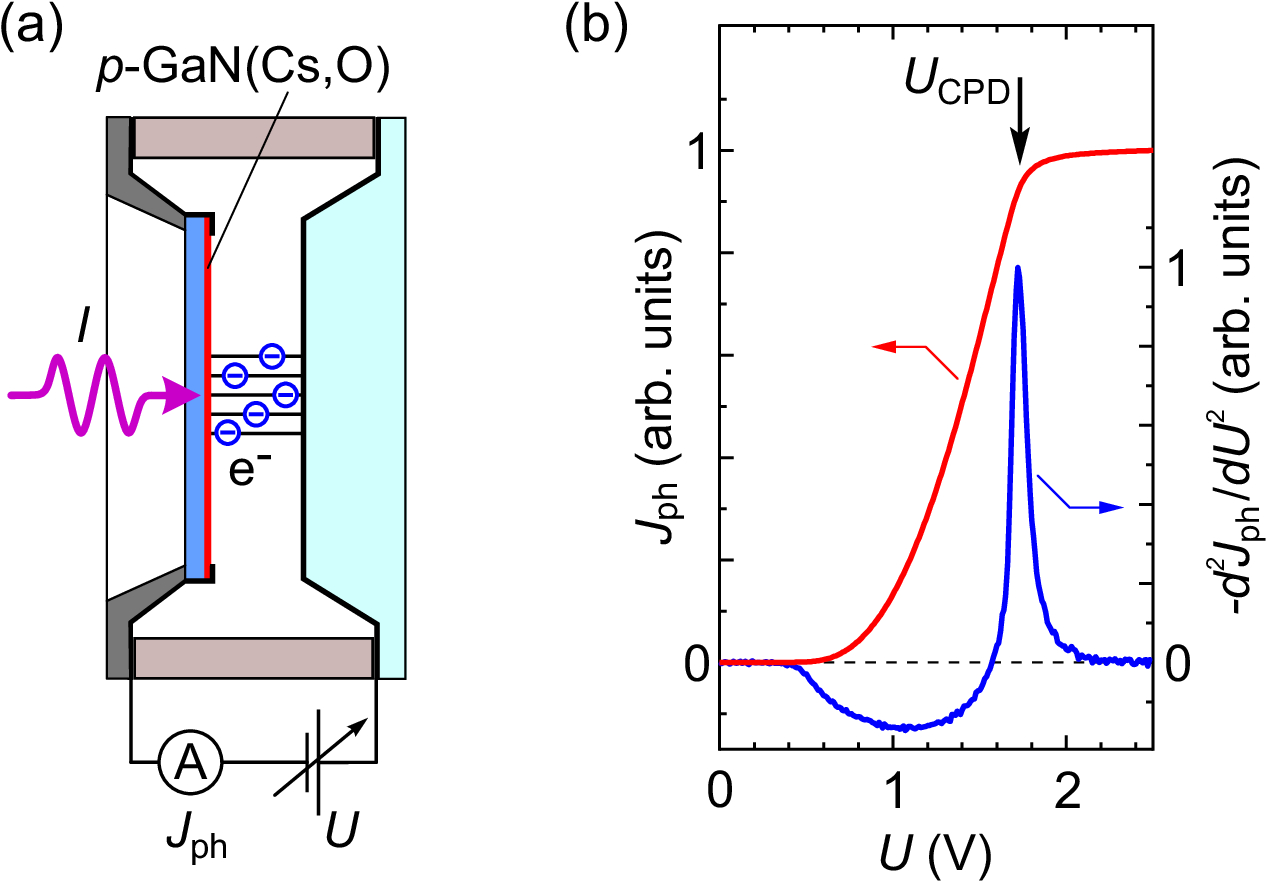}
\caption{\label{Fig.2} 
(a) Schematic cross section of a vacuum photodiode with a \textit{p}-GaN(Cs,O) NEA photocathode. $I$ is the optical power density, $J_{\text{ph}}$ is the photoemission current, $U$ is the external voltage. (b) Typical photocurrent-voltage characteristic of a photodiode and its second derivative. The contact potential difference of the illuminated vacuum photodiode $U_{\text{CPD}}$ is indicated with an arrow.
}
\end{figure}

\section{\label{sec:level1}Experimental Details}

In this study, the proposed method for determining the electron escape probability is applied to \textit{p}-GaN(Cs,O) NEA photocathodes \cite{Uchiyama2005, Pakhnevich2004, Sumiya2010, Rozhkov2018, Levenson2024, Sauty2024}. The AlN/GaN heterostructures were epitaxially grown by the metal-organic chemical vapour deposition on $\text{Al}_2\text{O}_3(0001)$ substrates. The thicknesses of the wurtzite \textit{p}-GaN layers were approximately 100\,nm. The concentrations of Mg and free holes at room temperature in the active \textit{p}-GaN layer were approximately $10^{19}~\text{cm}^{-3}$ and $10^{17}~\text{cm}^{-3}$, respectively. The surface preparation consisted in the removal of gallium oxides in the solution of HCl in isopropyl alcohol in a pure nitrogen atmosphere, air-tight transfer of samples to the ultra-high vacuum chamber and annealing in vacuum at approximately $600^{\circ}\text{C}$. Cesium and oxygen were deposited on a clean GaN surface at room temperature until the maximal QE was obtained. The details of the GaN surface preparation and activation with Cs and O$_2$ are described in Ref.~\cite{Tereshchenko2004}. 

The experiments were performed in vacuum photodiodes [Fig.~\ref{Fig.2}(a)] with semitransparent \textit{p}-GaN(Cs,O) photocathodes and metal anodes mounted parallel to each other in a hermetically sealed alumina ceramic bodies \cite{Rozhkov2018}. The working diameter of the photocathodes was equal to 18\,mm. The photocathode-anode gap varied from 0.3 to 3\,mm for different photodiodes. A halogen lamp with a diffraction grating monochromator was used to form a square-shaped light spot of approximately $5 \times 5~\text{mm}$ on the photocathodes. The optical power was varied by a set of calibrated neutral density filters. All measurements were performed at room temperature under illumination with photon energy $\hbar\omega = 3.65$\,eV exceeding the band gap of GaN $E_{\text{g}}^{\text{GaN}} \approx 3.4~\text{eV}$ \cite{Uchiyama2005}. The photocurrent was measured by a transimpedance amplifier followed by a low-pass filter with a time constant of $60$\,ms. The photocurrent variations due to the finite response time of our setup and transient currents due to the charging of the photodiode capacitance became smaller than the random noise within an interval of approximately 1\,s; therefore, we analyzed the photocurrent kinetics excluding this initial transient interval. The photovoltage variations were obtained from the evolution of the contact potential difference between the photocathode and anode $U_{\text{CPD}}$ on the optical power density. The $U_{\text{CPD}}$ was determined from the extremum in the second derivative of the photocurrent-voltage characteristic $J_{\text{ph}}(U)$ \cite{Rozhkov2018}, as is shown in Fig.~\ref{Fig.2}(b).

\section{\label{sec:level1}Results}

\begin{figure*}[t]
\includegraphics[width=0.88\linewidth]{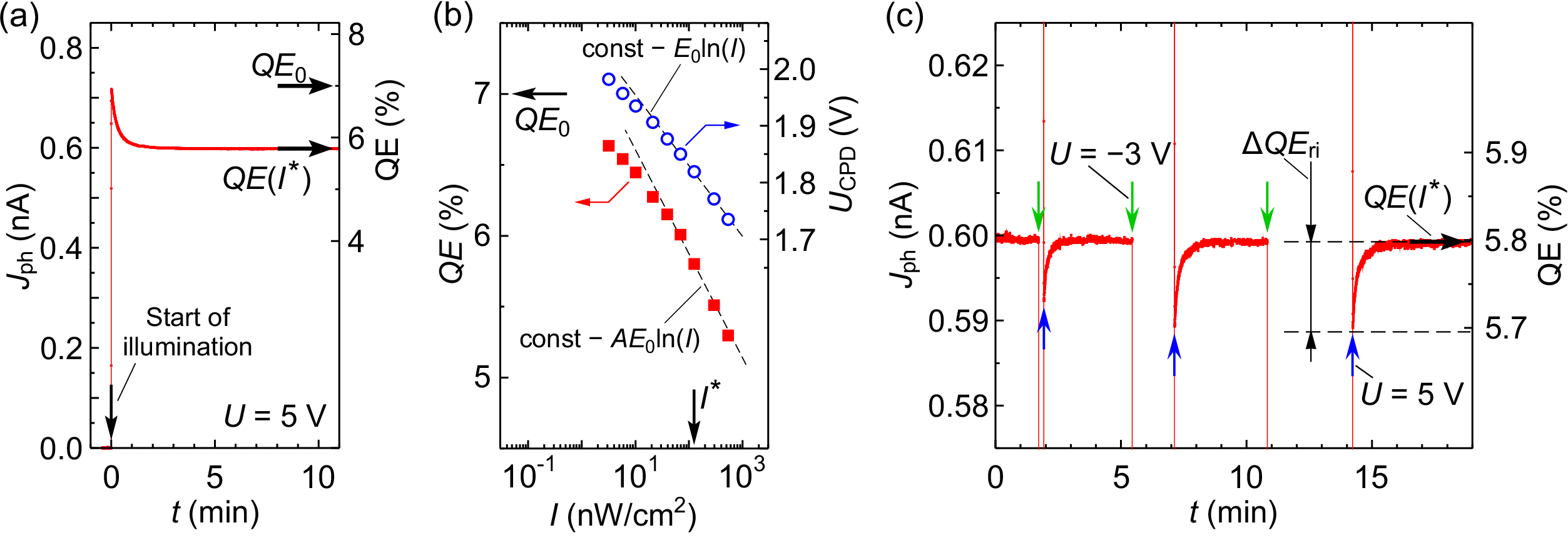}
\caption{\label{Fig.3} 
(a) The relaxational kinetics of the photoemission current $J_{\text{ph}}$ during the above-band-gap illumination of the \textit{p}-GaN(Cs,O) photocathode with optical power density $I^{*} = 120~\text{nW/cm}^{2}$. (b) Photoemission quantum efficiency (squares) and contact potential difference of the vacuum photodiode $U_{\text{CPD}}$ (circles) \textit{versus} optical power density. (c) The relaxational kinetics of photoemission current after keeping the illuminated photocathode in the retarding electric field. Switching to the retarding and accelerating electric fields occurs at the moments of time indicated by the downward and upward arrows, respectively. A QE decrease $\Delta QE_{\text{ri}}$ relative to the stationary value $QE(I^{*})$ is caused by the increase in SPV due to the reverse injection of photoemitted electrons in the retarding electric field .
}
\end{figure*}

The relaxational kinetics of the photoemission current $J_{\text{ph}}(t)$ upon the illumination of the \textit{p}-GaN(Cs,O) photocathode with optical power density $I^{*} = 120~\text{nW/cm}^{2}$ is shown in Fig.~\ref{Fig.3}(a). The measurements were performed on the vacuum photodiode with the photocathode-anode gap of 0.31\,mm. The photocathode was kept in the dark before the measurements to ensure the SPV relaxation; so, the photocurrent value at the start of illumination ($t = 0$\,min) is proportional to $QE_{0}$. As is seen, the illumination results in the photocurrent and, consequently, QE decrease by a factor of $(1.20 \pm 0.01)$ due to the SPV increase toward its stationary value $V_{\text{acc}}(I^{*})$.

As is seen in Fig.~\ref{Fig.3}(b), the dependences of the QE and $U_{\text{CPD}}$ on the optical power density $I$ are similar: at a high optical power limit, both QE and $U_{\text{CPD}}$ decrease as a natural logarithm of $I$, in agreement with Eqs.~\ref{eq.4} and ~\ref{eq.9}. This confirms that the QE decrease in Figs.~\ref{Fig.3}(a) and ~\ref{Fig.3}(b) is due to the increase in SPV. Fitting the slopes of the logarithmic dependences of QE and $U_{\text{CPD}}$ yields $AE_{0} = (0.32 \pm 0.03)\%$ and $E_{0} = (0.055 \pm 0.005)$\,V, respectively.

\begin{figure}[b]
\includegraphics[width=0.95\linewidth]{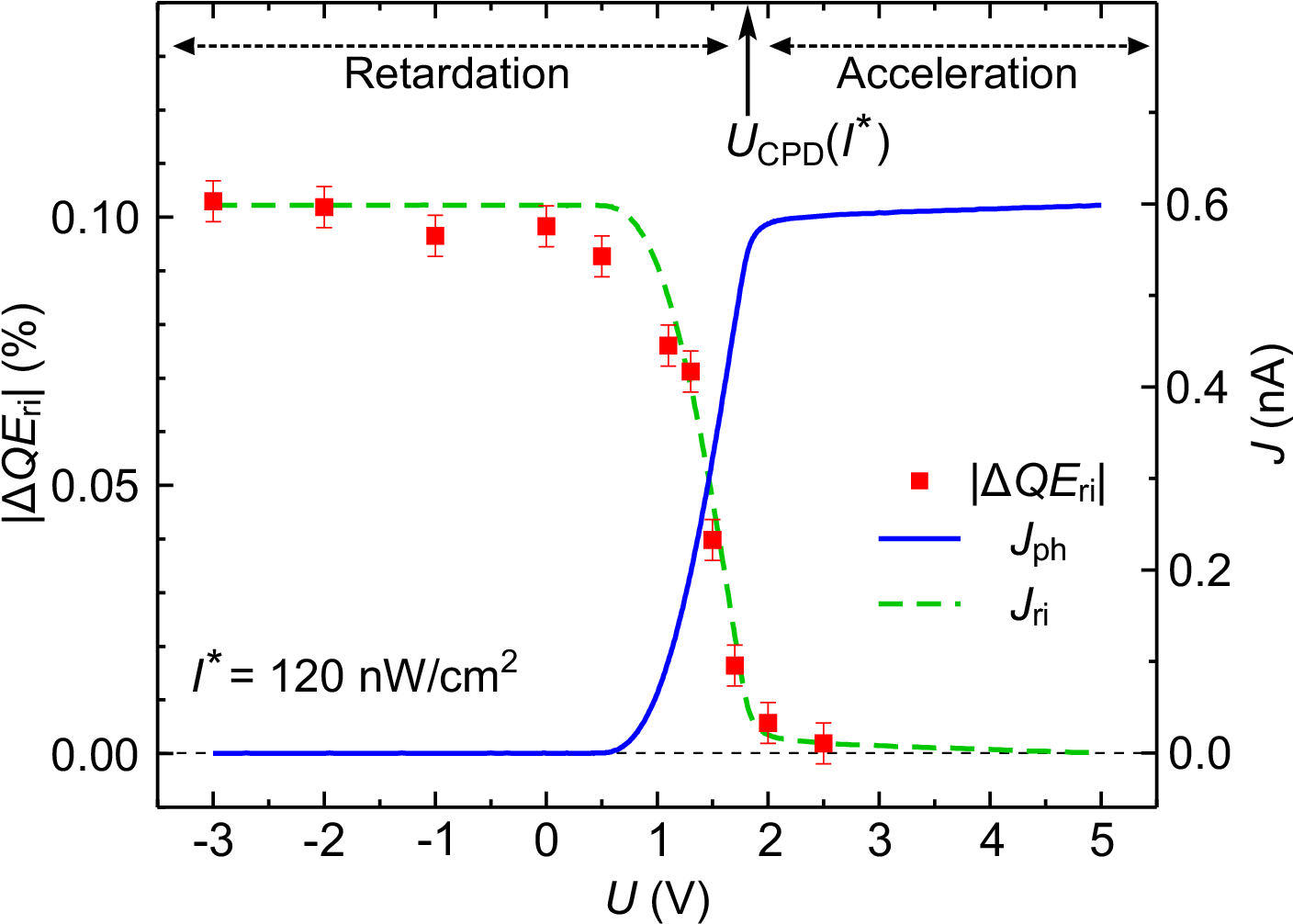}
\caption{\label{Fig.4} 
Modulus of reverse-injection-induced decrease of quantum efficiency $\Delta QE_{\text{ri}}$ (squares), photocurrent $J_{\text{ph}}$ (solid line) and reverse-injected current $J_{\text{ri}}$ (dashed line) \textit{versus} external voltage $U$ measured on the vacuum photodiode with the \textit{p}-GaN(Cs,O) photocathode at $I^{*} = 120~\text{nW/cm}^{2}$. The contact potential difference $U_{\text{CPD}}$ is indicated with an arrow.
}
\end{figure}

The relaxational kinetics of photoemission current $J_{\text{ph}}(t)$ after keeping the illuminated \textit{p}-GaN(Cs,O) photocathode in the retarding electric field is shown in Fig.~\ref{Fig.3}(c). It is seen that, after resuming the photocurrent measurements by applying the accelerating field, the initial values of the photocurrent and, consequently, QE are lower than the stationary values. The magnitude of the QE decrease saturates after keeping the photocathode in the retarding field for about 2\,min. The saturated value of $\Delta QE_{\text{ri}}$ equals to $(-0.103 \pm 0.003)\%$ for optical power density $I^{*}$.

In Fig.~\ref{Fig.4} we compared the dependence of $\Delta QE_{\text{ri}}$ modulus on the external voltage $U$ with the photocurrent-voltage characteristic of photodiode $J_{\text{ph}}(U)$, as well as with the dependence of the reverse-injected current on voltage $J_{\text{ri}}(U) = J_{\text{ph}}^{\text{max}} - J_{\text{ph}}(U)$, where $J_{\text{ph}}^{\text{max}}$ is the saturated value of the photocurrent. As is seen in  Fig.~\ref{Fig.4}, the $|\Delta QE_{\text{ri}}|$ closely follows the magnitude of reverse-injected current $J_{\text{ri}}$, which confirms that the quantum efficiency decrease $\Delta QE_{\text{ri}}$ is caused by the reverse injection of photoemitted electrons.

Using Eq.~\ref{eq.11} and the measured values of $\Delta QE_{\text{ri}}$ and $AE_{0}$, we obtain $P_{\text{e}}(I^{*}) = (27 \pm 3)\%$, that is, the escape probability of the photocathode at the photovoltage $V_{\text{acc}}(I^{*})$. The escape probability in the low optical power limit $P_{\text{e, 0}}$ is obtained by multiplying $P_{\text{e}}(I^{*})$ by the ratio $QE_{0}/QE(I^{*})$, which equals $(1.20 \pm 0.01)$ [see Fig.~\ref{Fig.3}(a)]. Thus, $P_{\text{e, 0}} = (32 \pm 3)\%$. 

Along with measuring the kinetics of QE according to the technique proposed here, for this particular photocathode we also determined SPV variations, after keeping photocathode in the retarding and accelerating fields, by measuring the CPD extracted from the photocurrent-voltage characteristics $J_{\text{ph}}(U)$ (a conventional technique), and obtained $P_{\text{e, 0}} = (40 \pm 10)\%$. Thus, although the conventional technique yielded less detailed and precise data due to relatively slow measuring the photocurrent-voltage characteristics, the resulting SPV kinetics and $P_{\text{e}}$ values determined by both techniques coincided within the experimental uncertainties. This coincidence is an additional verification of the proposed method. We found that the escape probabilities of various \textit{p}-GaN(Cs,O) photocathodes studied in this work lie in the range of 30--60\%. This agrees with the reported values of the electron escape probability in \textit{p}-GaN(Cs,O) \cite{Uchiyama2005, Sumiya2010}.

\section{\label{sec:level1}Discussion}

The proposed method is promising for further studying the dependence of the electron escape probability into vacuum on the structural and electronic properties of the emitting surfaces in semiconductor NEA photocathodes. A relatively broad range of $P_{\text{e}}$ values observed in this work in \textit{p}-GaN(Cs,O) photocathodes stresses the necessity of such studies. In this context, it is useful to discuss the capabilities and limitations of the new method.

First, the correct determination of $P_{\text{e}}$ using the proposed method requires that, at the retarding external electric field, all photoelectrons are injected back into the illuminated region of the photocathode. Therefore, the mean transverse displacement of photoelectrons between the points of photoemission and reverse injection, which is inversely proportional to the retarding electric field, must be significantly smaller than the light spot size. The fulfillment of this condition should be verified by measuring the dependence of $\Delta V_{\text{ri}}$ or $\Delta QE_{\text{ri}}$ on the retarding field magnitude. As can be seen in Fig.~\ref{Fig.4}, $\Delta QE_{\text{ri}}$ saturated at $U < 0$\,V for this particular photodiode with the photocathode-anode gap of 0.31\,mm. This indicates that practically all photoelectrons are injected back into the illuminated region of the \textit{p}-GaN(Cs,O) photocathode by a retarding field of about 100\,V/cm.

Second, a possible influence of anode charging on the photocurrent should be excluded, so that the photocurrent variations with increasing light intensity are caused only by the photocathode surface charging, which induce SPV. To meet this requirement, an anode should have sufficient electrical conductivity, i.e. the anode should be metallic, without oxide layers.

Third, as SPV decreases with increasing doping \cite{Rhoderick1978}, for high doping levels it can become too small for sufficiently accurate determination of the escape probability. This problem can be circumvented by lowering the photocathode temperature, which reduces the restoring current and, thus, increases SPV.

Finally, in the present work the proposed method is demonstrated for the transmission-mode photocathodes sealed in compact vacuum photodiodes with parallel-plate geometry, which provides well-defined electron trajectories. However, this method can be also explored in large vacuum set-ups and for the reflection-mode photocathodes, provided that the above mentioned requirement on the electron collection and reverse injection are met. It should be noted that the reflection-mode photocathodes are often illuminated through a metal grid, which allows one to collect electrons and apply electric fields \cite{Feng2015}. Care should be taken to account for a possible influence of the respective non-uniform illumination. Indeed, a fraction of electrons emitted from higher-intensity illuminated region may be reverse-injected into neighboring regions with lower intensity, thus reducing the obtained values of $P_{\text{e}}$.

\section{\label{sec:level1}Conclusion}

In summary, we proposed and developed a direct method for measuring the electron escape probability in semiconductor NEA photocathodes. The method does not require the data on the light absorption and electron transport in the bulk of the photocathode and is based on the variations in the surface photovoltage upon the reverse injection of emitted photoelectrons. We implemented the method by measuring the relaxational kinetics of the photoemission quantum efficiency when the polarity of the anode voltage changes from negative to positive, which blocks or extracts the photoemission current, respectively. This technique can be used in compact vacuum tubes, as well as in large vacuum setups. We applied the developed method for determining the electron escape probability in \textit{p}-GaN(Cs,O) photocathodes. The obtained results open up opportunities for optimizing the existing and developing novel semiconductor photocathodes.

\begin{acknowledgments}

The authors acknowledge the support from the Russian Science Foundation (Grant No.\,23-72-30003).

\end{acknowledgments}

\section*{Data Availability}

The data that support the findings of this study are available from the corresponding author upon reasonable request.

\providecommand{\noopsort}[1]{}\providecommand{\singleletter}[1]{#1}%

\end{document}